\documentstyle[aps]{revtex}
\begin{document}
\title{LANDAU-KHALATNIKOV-FRADKIN TRANSFORMATIONS AND THE FERMION
PROPAGATOR IN QUANTUM ELECTRODYNAMICS}
\author{A. Bashir and A. Raya} 
\address{Instituto de F{\'\i}sica y Matem\'aticas, 
         Universidad Michoacana de San Nicol\'as de Hidalgo\\
         Apartado Postal 2-82, Morelia, Michoac\'an 58040, M\'exico.}
\maketitle
\begin{abstract} 

  We study the gauge covariance of the massive fermion propagator 
in three as well as four dimensional Quantum Electrodynamics (QED). 
Starting from its 
value at the lowest order in perturbation theory, we evaluate
a non-perturbative expression for it by means of its 
Landau-Khalatnikov-Fradkin (LKF) transformation. We compare the
perturbative expansion of our findings with the known one loop results
and observe perfect agreement upto a gauge parameter independent term,
a difference permitted by the structure of the LKF transformations.

\vskip 0.5cm

\noindent
PACS number(s): 11.15.Tk,12.20.-m \hspace{6.8cm} UMSNH/02-8 

\end{abstract}

\section{Introduction}

         In a gauge field theory, Green functions transform in a specific 
manner under a variation of gauge. In Quantum Electrodynamics (QED)
these transformations carry the
name Landau-Khalatnikov-Fradkin (LKF) transformations, \cite{LK1,LK2,F1}.
Later,  they were derived  again by Johnson and Zumino through functional
methods,  \cite{JZ1,Z1}. These transformations are non-perturbative in
nature and hence have the potential of playing an important role in 
addressing the problems of gauge invariance which plague the strong 
coupling studies of Schwinger-Dyson equations (SDE). In general, the 
rules governing these transformations are far from simple. The fact 
that they are written in coordinate space adds to their complexity. As 
a result, these transformations have played less significant and
practical role in the study of SDE than desired. 
A consequence of gauge covariance are Ward-Green-Takahashi
identities (WGTI), \cite{W1,G1,T1}, which are simpler to use and,
therefore, have been extensively implemented in the above-mentioned studies.

 The LKF transformation for the three-point vertex is complicated and
hampers direct extraction of analytical restrictions on its
structure. Burden and Roberts, \cite{BR1}, carried out a numerical 
analysis to compare the self-consistency of various {\em ansatze} for
the vertex, 
\cite{BC1,H1,CP1}, by means of its LKF transformation.
In addition to these numerical constraints, indirect analytical
insight can be obtained on the non perturbative structure of the
vertex by demanding correct gauge covariance properties of the 
fermion propagator. There are numerous works in literature, based upon 
this idea, \cite{CP1,dong,BP1,BP2,BKP1,BT1} \footnote{A criticism of
the vertex construction in \cite{BT1} was raised in \cite{BKP2}}. 
However, all the work in this direction has been carried out in 
the context of massless QED3 and QED4. The masslessness of the
fermions implies that the fermion propagator can be written only in
terms of one function, the so called wavefunction renormalization, $F(p)$.
In order to
apply the LKF transform, one needs to know a Green function at least
in one particular gauge. This is a formidable task. However, one
can rely on approximations based on perturbation theory. It is
customary to take $F(p)=1$ in the Landau gauge, an
approximation justified by one loop calculation of the massless fermion
propagator in arbitrary dimensions, see for example, \cite{davydychev}.
The LKF transformation then implies a power law for $F(p)$ in QED4
and a simple trigonometric function in QED3. To improve upon these
results,
one can take two paths:
(i) incorporate the information contained in higher orders of
perturbation theory and (ii) study the massive theory. As pointed out
in \cite{BKP1}, in QED4, the power law structure of the wavefunction
renormalization remains intact by increasing order of
approximation in perturbation theory although the exponent of course
gets contribution from next to leading logarithms and so on 
\footnote{For the two loop calculation of the fermion propagator,
see for example \cite{Ross}}. In \cite{BKP1}, constraint was obtained
on the 3-point vertex by considering a power law where the exponent 
of this power law was not restricted only to the one loop fermion 
propagator. In QED3, the two loop fermion propagator was evaluated in
\cite{BKP2,BKP3,adnan1}, where it was explicitly shown that the 
the approximation $F(p)=1$ is only valid upto one loop, thus violating
the {\em transversality condition} advocated in \cite{BT1}. The result
found there was used used in \cite{adnan2} to find the improved
LKF transform. 

In the present article, we calculate the LKF transformed fermion propagator
in massive QED3 and QED4. We start with the simplest input which
corresponds to the lowest order of perturbation theory, i.e.,
$S(p)=1/ i {\not \! p}-m$ in the Landau gauge. On LKF transforming, 
we find the fermion propagator in an arbitrary covariant gauge.
In the case of QED3, we obtain the result in terms of basic functions
of momenta. In QED4, the final expression is in the form of
hypergeometric functions.
Coupling $\alpha$ enters as parameter of this transcendental
function. A comparison with perturbation theory needs the expansion
of the hypergeometric function in terms of its parameters. We use the 
technique developed by Moch {\em et. al.}, \cite{Moch}, for the said 
expansion.
We compare our results with the one loop expansion of the fermion
propagator in QED4 and QED3, \cite{Reenders,BR1}, and find perfect
agreement upto terms independent of the gauge parameter at one loop,
a difference permitted by the structure of the LKF transformations.
We believe that the incorporation of LKF transformations, along with
WGT identities, in the SDE can play a key role in addressing the 
problems of gauge invariance. For example, in the study of
the SDE of the fermion propagator, only those assumptions should be 
permissible which keep intact the correct behaviour of the Green
functions under the LKF transformations, in addition to ensuring
that the WGTI is satisfied. It makes it vital to explore how two and
three point Green functions transform in a gauge covariant fashion.
In this article, we consider only a two point function, namely, the
fermion propagator.

\section{Fermion Propagator and the LKF Transformation}

\noindent
We start by expanding out the fermion propagator, in momentum and 
coordinate spaces 
respectively, in its most general form as follows~:
\begin{eqnarray}
S_F(p;\xi)&=& A(p;\xi) + i  \frac{B(p;\xi)}{{\not \! p}}   \equiv
\frac{F(p;\xi)}{i {\not \! p}-{\cal M}(p;\xi)}
\;, 
\label{fpropmoment} \\
\nonumber   \\ S_F(x;\xi)&=&{\not \! x}X(x;\xi)+ Y(x;\xi) 
\label{fpropcoord} \;,
\end{eqnarray}
where $F(p;\xi)$ is generally referred to as the wavefunction
renormalization and ${\cal M}(p;\xi)$ as the mass function. 
$\xi$ is the usual covariant gauge parameter. Motivated from the
lowest order perturbation theory, we take
\begin{equation}
F(p;0)=1\hspace{.5cm}\mbox{and}\hspace{.5cm} {\cal M}(p;0)=m \;.
\label{lowestFM} 
\end{equation}
Perturbation theory also reveals that this result continues to hold true
to one loop order for the wavefunction renormalization.
Eqs.~(\ref{fpropmoment},\ref{fpropcoord}) are related to each
other through the following Fourier transforms~:
\begin{eqnarray}
S_F(p;\xi)&=&\int d^dx e^{i p\cdot x}S_F(x;\xi)   \label{pFourier} \\ 
S_F(x;\xi)&=&\int\frac{d^dp}{(2\pi )^d}e^{-i p\cdot x}S_F(p;\xi)
\;,
\label{xFourier}
\end{eqnarray}
where $d$ is the dimension of space-time. The LKF transformation
relating the coordinate space fermion propagator in Landau gauge
to the one in an arbitrary covariant gauge reads~:
\begin{equation}
S_F(x;\xi) = S_F(x;0){\rm e}^{-i [\Delta_d(0)-\Delta_d(x)]} \;,
\end{equation}
where
\begin{equation}
\Delta_d(x)=-i \xi e^2\mu^{4-d}\int_0^\infty\frac{d^dp}{(2\pi )^d}
\frac{e^{- i p\cdot x}}{p^4} \; .
\end{equation}
$e^2$ is the dimensionless electromagnetic coupling. 
Taking $\psi$ to be the angle between $x$ and $p$, we can write
$d^dp=dp p^{d-1}\sin^{d-2}\psi d\psi \Omega_{d-2}$, where 
$\Omega_{d-2}=2 \; \pi^{(d-1)/2}/\Gamma\left((d-1)/2 \right)$. Hence
\begin{equation}
\Delta_d(x)=-i \xi e^2\mu^{4-d}f(d)\int_0^\infty dp p^{d-5}
\int_0^{\pi} d\psi \sin^{d-2}\psi e^{-i p x \cos \psi} \;,
\end{equation}
where $f(d)=\Omega_{d-2}/(2\pi )^d$. Performing angular and radial 
integrations, we arrive at the following equation
\begin{equation}
\Delta_d (x)=-\frac{i \xi e^2}{16 (\pi)^{d/2}}
(\mu x)^{4-d}\Gamma\left(\frac{d}{2}-2\right) \;.\label{deltad}
\end{equation}
With these tools at hand, the procedure now is as follows~:
\begin{itemize}
\item Start with the lowest order fermion propagator and Fourier 
transform it to coordinate space.

\item Apply the LKF transformation law.

\item Fourier transform the result back to momentum space.

\end{itemize}

\section{Three dimensional case} 

\noindent
Employing
Eqs.~(\ref{fpropmoment},\ref{fpropcoord},\ref{lowestFM},\ref{xFourier}),
the lowest order three dimensional fermion propagator in Landau gauge
in the position space is given by
\begin{eqnarray}
X(x;0)&=& -\frac{e^{-m x}(1+ m x)}{4\pi x^3} \;,\\ \nonumber
Y(x;0)&=& -\frac{m e^{-m x}}{4\pi x}  \;.
\end{eqnarray}
Once in the coordinate space, we can apply the LKF transformation
law using expression (\ref{deltad}) explicitly in three dimensions~:
\begin{equation}
\Delta_3(x)=-\frac{i \alpha \xi x}{2} \;,
\end{equation}
where $\alpha=e^2/4 \pi$. The fermion propagator in an arbitrary gauge
is then
\begin{eqnarray}
S(x;\xi)=S_F(x;0)e^{-(\alpha\xi/2)x} \;.
\end{eqnarray}
For Fourier transforming back to momentum space, we use
\begin{eqnarray}
A(p;\xi)&=&-\frac{F(p;\xi ){\cal M}(p;\xi)}{p^2+{\cal M}^2(p;\xi )}
=\int d^3x \; e^{i p\cdot x} \; Y(x;\xi) \, \\ \nonumber
i B(p;\xi)&=&-\frac{i p^2F(p;\xi)}{p^2+{\cal M}^2(p;\xi)}
=\int d^3x \; p \cdot x \; e^{i p\cdot x} \; X(x;\xi) \;.
\end{eqnarray}
Performing the angular integration, we get
\begin{eqnarray}
A(p;\xi)&=& - \frac{m}{p} \; \int_0^{\infty} dx \; \sin{px} \,
{\rm e}^{-(m+\alpha \xi/2)x} 	\;, \\ \nonumber \\
 B(p;\xi)&=& \frac{1}{p} \; \int_0^{\infty} \frac{dx}{x^2} \, (1+mx) \,
\left[ px \cos px - \sin px \right] {\rm e}^{-(m+\alpha \xi/2)x} \;, 
\end{eqnarray}
and the radial integration then yields
\begin{eqnarray}
A(p;\xi)&=&-\frac{4 m}{4 p^2 + \left( 2 m+ {\alpha \xi} \right)^2 } \;\\ 
B(p;\xi)&=&- \frac{4p^2+\alpha\xi (2m+\alpha\xi)}{4p^2+(2m+\alpha\xi)^2}
+ \frac{\alpha\xi}{2p}\arctan{ \left[ 2 p/(2 m+ {\alpha\xi})
\right]} \;.
\end{eqnarray}
One can now arrive at the following expressions for the wavefunction 
renormalization and the mass function, respectively~:
\begin{eqnarray}
F(p;\xi)&=&-\frac{\alpha\xi}{2p} 
\arctan{\left[2p/(2m+\alpha\xi) \right]}+ 
\frac{2p(4p^2+\alpha^2\xi^2)-\alpha\xi (4p^2+\alpha\xi(2m+\alpha\xi))
\arctan{[2p/(2m+\alpha\xi)]}}{ 2p(4p^2+\alpha\xi(2m+\alpha\xi))-
\alpha\xi(4p^2+(2m+\alpha\xi)^2)\arctan{\left[2p/(2m+\alpha\xi) 
\right]}  }  \;,  \label{LKFF3} \\
{\cal M}(p;\xi)&=&\frac{8p^3 m}{2p(4p^2+\alpha\xi(2m+\alpha\xi))-
\alpha\xi(4p^2+(2m+\alpha\xi)^2)\arctan{\left[2p/(2m+\alpha\xi) 
\right]} } \;.   \label{LKFM3}
\end{eqnarray}
In the massless limit, one immediately recuperates the well-known results~:
\begin{eqnarray}
F_{\rm
massless}(p;\xi)&=&1-\frac{\alpha\xi}{2p}\arctan{\frac{2p}{\alpha\xi}} 
\;,
\\ \nonumber
{\cal M}_{\rm massless}(p;\xi)&=&0 \;.
\end{eqnarray}
In the weak coupling, we can expand out Eqs.(\ref{LKFF3},\ref{LKFM3})
in powers of $\alpha$. To ${\cal O}(\alpha)$, we find
\begin{eqnarray}
   F(p;\xi)&=&1+\frac{\alpha\xi}{2p^3}\left[(m^2-p^2) \; {\rm
arctan} \; [p/m] - m p\right] \;,   \\
 {\cal M}(p;\xi)&=&m\left[ 1+\frac{\alpha\xi}{2p^3}\left\{ (m^2+p^2) \; {\rm
arctan} \; [p/m] - m p \right\} \right] \;.    
\end{eqnarray}
The expression for the wavefunction renormalization function fully
matches with that obtained in ~\cite{BR1}, while the one for the mass 
function is also in agreement upto a term proportional to 
$\alpha\xi^0$, as allowed by the structure of the LKF transformations.

\section{Four dimensional case}

\noindent
Employing
Eqs.~(\ref{fpropmoment},\ref{fpropcoord},\ref{lowestFM},\ref{xFourier}),
the
lowest order four dimensional fermion propagator in position space
is given by
\begin{eqnarray}
X(x;0)&=&-\frac{m^2}{4\pi^2x^2} \;  K_2(mx)  \;,\\ 
Y(x;0)&=&-\frac{m^2}{4\pi^2x} \; K_1(mx) \;,
\end{eqnarray}
where $K_1$ and $K_2$ are Bessel functions of the second kind. In
order to apply the LKF transformation in four dimensions, we expand
Eq.~(\ref{deltad}) around $d=4- \epsilon$ and use the following identities
\begin{eqnarray*}
\Gamma\left(-\frac{\epsilon}{2}\right)&=&-\frac{2}{\epsilon}-\gamma
+{\cal O}(\epsilon)  \;, \\ \nonumber
x^{\epsilon}&=&1+\epsilon \ln{x}+{\cal O}(\epsilon^2) \;,
\end{eqnarray*}
to obtain
\begin{equation}
\Delta_4(x)=i \frac{\xi e^2}{16\pi^{2-\epsilon/2}}
\left[ \frac{2}{\epsilon}+\gamma+2\ln{\mu x}+{\cal O}(\epsilon)\right]
\; .
\end{equation}
Note that we cannot write a similar expression for $\Delta_4(0)$
because of the presence of the term proportional to
$\ln{x}$. Therefore, we introduce a cut-off scale $x_{min}$. Now
\begin{equation}
\Delta_4(x_{min})-\Delta_4(x)=-i \ln{\left(\frac{x^2}{x_{min}^2}\right)^\nu},
\end{equation}
where $\nu=\alpha\xi/4\pi$. Hence
\begin{equation}
S_F(x;\xi)=S_F(x;0)\left(\frac{x^2}{x_{min}^2}\right)^{-\nu} .
\end{equation}
For Fourier transforming back to momentum space we use the following 
expressions
\begin{eqnarray}
A(p;\xi)&=&-\frac{F(p;\xi ){\cal M}(p;\xi)}{p^2+{\cal M}^2(p;\xi )}
=\int d^4x \; e^{i p\cdot x} \; Y(x;\xi) \, \\ \nonumber
i B(p;\xi)&=&-\frac{i p^2F(p;\xi)}{p^2+{\cal M}^2(p;\xi)}
=\int d^4x \; p \cdot x \; e^{i p\cdot x}    \; X(x;\xi) \;.
\end{eqnarray}
On carrying out angular integration, we obtain~:
\begin{eqnarray}
A(p;\xi)&=& - \frac{m^2}{p} \, x_{min}^{2 \nu} \;
\int_0^{\infty} dx x^{-2 \nu+1} \; K_1(mx) \, J_1(px) \;, \\
B(p;\xi)&=& - m^2 \;
\int_0^{\infty} dx x^{-2 \nu+1} \; K_2(mx) \, J_2(px) \;.
\end{eqnarray}
The radial integration then yields~:
\begin{eqnarray}
A(p;\xi)&=&- \frac{1}{m} \; \left( \frac{m^2}{\Lambda^2} \right)^{\nu} 
\;
\Gamma(1-\nu)\Gamma(2-\nu) 
\; {~}_2F_1\left(1-\nu,2-\nu;2;-\frac{p^2}{m^2}\right)  \;, \\
\nonumber \\
B(p;\xi)&=& -\frac{p^2}{2 m^2} \; 
\left( \frac{m^2}{\Lambda^2} \right)^{\nu} \;
\Gamma(1-\nu)
\Gamma(3-\nu) \; {~}_2F_1\left(1-\nu,3-\nu;3;-\frac{p^2}{m^2}\right) \;,
\end{eqnarray}
where we have identified $2/x_{min} \rightarrow \Lambda$. The above
equations imply
\begin{eqnarray}
\nonumber \\ F(p;\xi)&=&
\frac{ \Gamma(1-\nu)}{
2 m^2 \; \Gamma(3-\nu) \;  {~}_2F_1\left(1-\nu,3-\nu;3;
-p^2/m^2 \right)} \; \left( \frac{m^2}{\Lambda^2} \right)^{\nu} \nonumber  \\ 
&\times&\Bigg[4m^2\Gamma^2(2-\nu) \; {~}_2F_1^2\left(1-\nu,2-\nu;2;
-\frac{p^2}{m^2} \right)+p^2 \; \Gamma^2(3-\nu) \; 
_2F_1^2\left( 1-\nu,3-\nu;3;-\frac{p^2}{m^2}\right) \Bigg] \;,
\label{ourresultF}  \\ \nonumber \\
{\cal M}(p;\xi)&=&\frac{2m \; {~}_2F_1\left(
1-\nu,2-\nu;2;- p^2/m^2 \right)} 
{(2-\nu) \; {~}_2F_1\left(1-\nu,3-\nu;3;- p^2/m^2 \right)} \;.
\label{ourresultM} 
\end{eqnarray}
Eqs.~(\ref{ourresultF},\ref{ourresultM}) constitute the LKF transformation
of Eqs.~(\ref{lowestFM}). We shall now see that although 
Eqs.~(\ref{lowestFM}) correspond to the lowest order propagator, their
LKF transformation, Eqs.~(\ref{ourresultF},\ref{ourresultM}), is 
non-perturbative in nature and contains information of higher orders.

\subsection{Case $\alpha=0$}

\noindent
Let us switch off the coupling and put $\alpha=0$ which implies
$\nu=0$. Now using the identity
\begin{eqnarray}
   _2F_1(1,2;2;-p^2/m^2)&=& \, _2F_1(1,3;3;-p^2/m^2)= (1+p^2/m^2)^{-1} \;,
\end{eqnarray}
it is easy to see that
\begin{equation}
F(p;\xi)=1\hspace{.5cm}\mbox{and}\hspace{.5cm} {\cal M}(p;\xi)=m \;,
\label{lowestFMxi} 
\end{equation}
which coincides with the lowest order perturbative result as expected.

\subsection{Case $m >> p$}

\noindent
In the limit $m >> p$, the hypergeometric functions in 
Eqs.~(\ref{ourresultF},\ref{ourresultM}) can be easily
expanded in powers of $p^2/m^2$, using the identity
\begin{equation}
_2F_1\left(\alpha,\beta;\gamma;-\frac{p^2}{m^2} \right)=1-
\frac{\alpha\beta}{\gamma} \frac{p^2}{m^2} +{\cal O} 
\left(\frac{p^2}{m^2} \right)^2 \; .
\end{equation}
Retaining only ${\cal O}(p^2/m^2)$ terms, we arrive at~:
\begin{eqnarray}
 F(p;\xi)&=& 
\frac{\Gamma(1-\nu) \Gamma(2-\nu)}{(1-\nu/2)} \; \left[ 
1 + \frac{2 \nu}{3} \; \left( 1 - \frac{5 \nu}{8} \right)
\frac{p^2}{m^2}\; + {\cal O} 
\left(\frac{p^2}{m^2} \right)^2 \right] 
\left( \frac{m^2}{\Lambda^2} \right)^{\nu}  ,  \\
{\cal M}(p;\xi)&=& \frac{m}{(1-\nu/2)} \left[ 1 + \frac{\nu}{6}
(1-\nu) \frac{p^2}{m^2} + {\cal O} 
\left(\frac{p^2}{m^2} \right)^2 \right] \;.
\end{eqnarray}
Now carrying out an expansion in $\alpha$ and substituting $\nu =
\alpha \xi/ 4 \pi$, we get the following ${\cal O}(\alpha)$ expressions~:
\begin{eqnarray}
F(p;\xi)&=&1+\frac{\alpha\xi}{4\pi}\left[2\gamma
-\frac{1}{2}+\frac{2p^2}{3m^2}+\ln{\frac{m^2}{\Lambda^2}}\right] \;, 
\label{largemF}
\\ 
{\cal
M}(p;\xi)&=&m\left\{1+\frac{\alpha\xi}{8\pi}
\left[1+\frac{p^2}{3m^2}\right]\right\}  \;.  \label{largemM}
\end{eqnarray}
Let us now compare these expressions against the one-loop perturbative
evaluation of the massive fermion propagator, see e.g., \cite{Reenders}~:
\begin{eqnarray}
F_{\rm 1-loop}(p;\xi)&=&1-\frac{\alpha\xi}{4\pi}
\left[ C\mu^\epsilon+\left(1-\frac{m^2}{p^2}\right) (1-L)\right] \;, 
 \label{KRPF} \\  
{\cal M}_{\rm 1-loop}(p;\xi)&=&m+\frac{\alpha m}{\pi}
\left[\left(1+\frac{\xi}{4}\right)+\frac{3}{4}(C\mu^\epsilon-L)
+\frac{\xi}{4}\frac{m^2}{p^2}(1-L)\right] \label{KRPMC} \;,
\end{eqnarray}
where
\begin{eqnarray}
L&=&\left(1+\frac{m^2}{p^2} \right) \ln{\left(1+\frac{p^2}{m^2}\right)}
\;,
\nonumber \\ \nonumber
C&=&-\frac{2}{\epsilon}-\gamma-\ln{\pi}-\ln{\left(\frac{m^2}{\mu^2}\right)}
\;.
\end{eqnarray}
          Knowing the fermion propagator even in one particular gauge is a
prohibitively difficult task. Therefore, Eqs.~(\ref{lowestFM}) have to
be viewed only as an approximation. For the wavefunction
renormalization $F(p)$, this approximation is valid upto one loop
order, whereas, for the mass function, it is true only to the lowest
order. Therefore we cannot expect the LKF transform of
Eqs.~(\ref{lowestFM}) to yield correctly each term in the perturbative
expansion of the fermion propagator. However, it should correctly
reproduce all those terms at every order of expansion which vanish in
the Landau gauge at ${\cal O}(\alpha)$ and beyond. Therefore, we
expect Eq.~(\ref{KRPF}) to be exactly reproduced and Eq.~(\ref{KRPMC}) to be 
reproduced upto the terms which vanish in the Landau gauge at 
${\cal O}(\alpha)$. After subtracting these terms, we call the
resulting function as the {\em subtracted} mass function~:
\begin{equation}
{\cal M}_{\rm 1-loop}^S(p;\xi)=
m+\frac{\alpha\xi m}{4\pi}\left[1+\frac{m^2}{p^2}(1-L)\right] \; .
\label{KRPMS}
\end{equation}
In the limit $m\rightarrow\infty$, the wavefunction renormalization 
acquires the form
\begin{equation}
F(p;\xi)_{\rm 1-loop} =1+\frac{\alpha\xi}{4\pi}
\left[- C\mu^\epsilon-\frac{1}{2}+\frac{2p^2}{3m^2}\right] \;,
\end{equation}
while the {\em subtracted} mass function is
\begin{equation}
{\cal M}_{\rm 1-loop}^S(p;\xi)= 
m\left\{1+\frac{\alpha\xi}{8\pi}\left[1+\frac{p^2}{3m^2}\right]\right\}
\:.
\end{equation}
The last two expressions are in perfect agreement with 
Eqs.~(\ref{largemF},\ref{largemM}) after we make the identification~:
\begin{equation}
- C\mu^\epsilon \rightarrow  2\gamma + \ln{\frac{m^2}{\Lambda^2}} \;.
\label{cutoffdimreg}
\end{equation}

\subsection{Case of weak coupling} 

\noindent
The case $m >> p$ is relatively easier to handle as we merely have to expand
$_2F_1(\beta, \gamma; \delta; x)$ in powers of $x$ and retain only the
leading terms. If we want to obtain a series in powers of the coupling
alone, we need the expansion of the hypergeometric functions 
in terms of its parameters $\beta$ and $\gamma$. We follow the
technique developed in ~\cite{Moch}. One of the mathematical objects
we shall use for such an expansion are the $Z$-sums defined as~:
\begin{equation}
Z(n;m_1, \ldots ,m_k;x_1, \ldots , x_k)=
\sum_{n\ge i_1>i_2>\ldots >i_k>0} \frac{x_1^{i_1}}{i_1^{m_1}}\ldots 
\frac{x_k^{i_k}}{i_k^{m_k}} \;.   
\end{equation}
For $x_1=\ldots =x_k=1$ the definition reduces to the Euler-Zagier
sums, \cite{Zsums1,Zsums2}~:
\begin{equation}
Z(n; m_1, \ldots ,m_k;1, \ldots ,1)=Z_{m_1,\ldots ,m_k}(n) \;.
\end{equation}
Euler-Zagier sums can be used in the expansion of Gamma functions. For 
positive integers $n$ we have~\cite{Moch}:
\begin{equation}
\Gamma(n+\epsilon)=\Gamma (1+\epsilon)\Gamma (n) \left[ 
1+\epsilon Z_1(n-1)+\ldots +
\epsilon^{n-1}Z_{11\ldots 1}(n-1) \right] \;. \label{gammaexp}
\end{equation}
The first sum $Z_1(n-1)$, e.g., is just the $(n-1)$-th harmonic number, 
$H_{n-1}$, of order 1~:
\begin{equation}
Z_1(n-1)=\sum_{i=1}^{n-1}\frac{1}{i} \equiv H_{n-1}  \;.
\end{equation}
With these definitions in hand, we proceed to expand a hypergeometric
function, $_2F_1(1+\varepsilon,2+\varepsilon;2;x)$, as an example, 
assuming $\vert x \vert<1$~:
\begin{eqnarray}
\nonumber _2F_1(1+\varepsilon,2+\varepsilon;2;x)&=&1+
\frac{\Gamma (2)}{\Gamma (1+\varepsilon)
\Gamma(2+\varepsilon)}\sum_{n=1}^\infty 
\frac{\Gamma(1+\varepsilon+n)\Gamma(2+\varepsilon+n)}{\Gamma(2+n)} 
\frac{x^n}{n!}\\ \nonumber
&=&1+\frac{1}{(1+\varepsilon)\Gamma^2(1+\varepsilon)}
\sum_{n=1}^\infty \frac{(1+\varepsilon+n)(\varepsilon+n)^2
\Gamma^2(\epsilon+n)}{\Gamma(2+n)} \frac{x^n}{n!} \;.
\end{eqnarray}
Employing Eq.~(\ref{gammaexp}), we can expand the last expression 
in powers of $\epsilon$ to any desired order of approximation. We
shall be interested only in terms upto ${\cal O}(\alpha)$.
\begin{eqnarray}
_2F_1(1+\varepsilon,2+\varepsilon;2;x)
&=&1+\sum_{n=1}^\infty x^n-\varepsilon\sum_{n=1}^\infty
x^n+\varepsilon
\sum_{n=1}^\infty \frac{2+3n}{n(n+1)}x^n+2\varepsilon \sum_{n=1}^\infty
H_{n-1}x^n \;.
\end{eqnarray}
Performing the summations, we obtain
\begin{equation}
_2F_1(1+\varepsilon,2+\varepsilon;2;x)=\frac{1}{1-x}
\left[ 1-\varepsilon\left\{
1+ \frac{1+x}{x} \; \ln{(1-x)}\right\} \right] \;.
\end{equation}
Similarly,
\begin{equation}
_2F_1(1+\varepsilon,3+\varepsilon;3;x)=\frac{1}{1-x}-
\varepsilon\left\{ \frac{1}{x}+\frac{3}{2}\frac{1}{1-x}+
\left(\frac{1+x}{x^2}+\frac{2}{1-x} \right)
\ln{(1-x)}\right\} \;.
\end{equation}
Substituting back into Eqs.~(\ref{ourresultF},\ref{ourresultM}) and 
identifying $\varepsilon=-\nu$, we obtain 
\begin{eqnarray}
F(p;\xi)&=&1-\frac{\alpha\xi}{4\pi}
\left[-2\gamma-\ln{\frac{m^2}{\Lambda^2}}+
\left(1-\frac{m^2}{p^2}\right)(1-L)\right]  \;, \\ \nonumber
{\cal M}(p;\xi)&=&m+\frac{\alpha\xi m}{4\pi}\left[1+\frac{m^2}{p^2}
(1-L)\right]  \;,
\end{eqnarray}
which matches exactly onto the one loop result of 
Eqs.~(\ref{KRPF},\ref{KRPMS}) after the same identification as before,
i.e., (\ref{cutoffdimreg}). Therefore, we have seen that the LKF
transformation of the bare propagator contains important information
of higher orders in perturbation theory.

\section{Conclusions}
We have studied the gauge covariance of the massive fermion propagator
in three as well as four dimensional QED through its LKF
transformation, starting from its lowest order approximation.
Eqs.~(\ref{LKFF3},\ref{LKFM3},\ref{ourresultF},\ref{ourresultM}) form
the main result of this article. In the three dimensional case, the
LKF transformation consists of basic functions of the momentum variable,
whereas, in the four dimensional case, hypergeometric functions arise
with electromagnetic coupling as parameter of these functions.
Although our input is only the bare propagator, the corresponding
LKF transformation, being non-perturbative in nature, contains useful
information of higher orders in perturbation theory. For example, we 
have shown that a perturbative expansion of our results matches onto
the known 1-loop results upto gauge independent terms at this order.
This slight difference arises due to our approximated input and can 
be corrected systematically at the cost of increasing complexity of
the integrals involved. We intend to carry out similar exercise for
the 3-point fermion-boson vertex. LKF transformations of the
propagator and the vertex impose useful constraints on the SDE and we 
believe that these transformations can be of immense help in
addressing the problems of gauge invariance in the related studies.

\section*{Acknowledgments}

\noindent
We are grateful to C. Schubert for bringing to our attention article 
\cite{Moch}. We thank CIC and Conacyt for their support under 
grants 4.10 and 32395-E respectively. AB also acknowledges the
financial support by Sistema Nacional de Investigadores (SNI). 

\section*{Appendix}

Most of the integrals involved in this paper are listed below for a
quick reference~\cite{tables,moretables}~:

\begin{eqnarray}
\int_0^\pi d\psi \sin^{d-2}\psi\cos{\psi}e^{-\iota p x
\cos{\psi}}&=&- i \sqrt{\pi}\left(\frac{px}{2}\right)^{1-\frac{d}{2}}
\Gamma\left(\frac{d-1}{2}\right)J_{\frac{d}{2}}(px) \;,  \\  \nonumber \\
\int_0^{\infty} x^{d/2-1} \; J_{d/2}(ax) &=& 
\frac{\Gamma(d/2)}{2^{1-d/2} \; a^{d/2}}  \;.
\end{eqnarray}
For the three dimensional case, the needed integrals are~:
\begin{eqnarray}
\int_0^\pi d\theta \; \sin{\theta} \; e^{-i px\cos{\theta}}
&=&\frac{2\sin{px}}{px}   \;, \\ \nonumber \\
\int_0^\pi d\theta \; \cos{\theta} \; \sin{\theta} \; e^{- i
px\cos{\theta}}&=& 
2 i \left[ \frac{\cos{px}}{px }-\frac{\sin{px}}{(px)^2} \right] \;,\\
\nonumber \\
\int_0^\infty dp  \;
\frac{p^3}{(p^2+m^2)}   \left[ \frac{\cos{px}}{px}
-\frac{\sin{px}}{(px)^2}\right]
&=& -\frac{\pi}{2} \; \frac{(1+mx)}{x^2} \; e^{-mx} \;, \\ \nonumber \\
\int_0^\infty dp\frac{p\sin{px}}{(p^2+m^2)}&=&\frac{\pi}{2}e^{-mx} \;,
\\
\nonumber \\
\frac{1}{p}\int_0^\infty\frac{dx}{x^2}e^{-ax}[px\cos{px}-\sin{px}]&=&-1
+\frac{a}{p}\arctan{\frac{p}{a}} \;, \\
\nonumber \\
\frac{1}{p}\int_0^\infty\frac{dx}{x}e^{-ax}[px\cos{px}-\sin{px}]&=& 
\frac{a}{a^2+p^2} - \frac{1}{p}\arctan{\frac{p}{a}} \;, \\
\nonumber \\
\int_0^\infty dx\sin{px}e^{-(m+ \alpha\xi/2)x}&=&
\frac{p}{\left(m+ \alpha\xi/2 \right)^2+p^2}  \;.
\end{eqnarray}
For the four dimensional case, we used the following integrals in particular~:
\begin{eqnarray}
\int_0^{\pi} d \theta {\sin}^2 \theta e^{- i px \cos \theta} &=&
\frac{\pi}{p x} \; J_1(px)  \;, \\ \nonumber \\ 
\int_0^\infty dp \frac{p^{\nu+1}J_\nu(px)}{(p^2+m^2)^{\mu+1}}&=& 
\frac{m^{\nu-\mu}x^\mu}{2^\mu\Gamma(\mu+1)}K_{\nu-\mu}(mx) \;, \\ \nonumber \\
\int_0^\infty dx x^{-\lambda}K_\mu (ax)J_\nu (bx)&=& 
\frac{a^{\lambda-\nu-1}  b^\nu}{2^{\lambda+1}\Gamma(1+\nu)}
  \Gamma\left(\frac{\nu-\lambda+\mu+1}{2}\right)
\Gamma\left(\frac{\nu-\lambda-\mu+1}{2}\right)  \nonumber \\
&& \hspace{15mm} \times {~}_2F_1
\left(\frac{\nu-\lambda+\mu+1}{2},
\frac{\nu-\lambda-\mu+1}{2};\nu+1;-\frac{b^2}{a^2}\right) \;.
\end{eqnarray}
Some of the series used in our calculation are as follows~:
\begin{eqnarray}
\sum_{n=1}^{\infty}  H_{n-1} x^{n} &=& - \frac{x \ln (1-x)}{1-x} \;, \\
\nonumber \\
\sum_{n=1}^{\infty} \frac{n+1}{n (n+2)} \; x^n &=& - \frac{2+x}{4x} -
\frac{(1+x^2) \ln(1-x)}{2 x^2}  \;, \\ \nonumber \\
\sum_{n=1}^{\infty} \frac{1}{(n+1)(n+2)} \; x^{n} &=& \frac{2-x}{2x} +
\frac{(1-x) \ln(1-x)}{x^2} \;.
\end{eqnarray}


\begin{thebibliography}{55}
\bibitem{LK1} L.D. Landau and I.M. Khalatnikov, Zh. Eksp. Teor. Fiz. {\bf 29} 
89 (1956).
\bibitem{LK2} L.D. Landau and I.M. Khalatnikov, Sov. Phys. JETP {\bf 2} 69
(1956). 
\bibitem{F1} E.S. Fradkin, Sov. Phys. JETP {\bf 2} 361 (1956).
\bibitem{JZ1} K. Johnson and B. Zumino, Phys. Rev. Lett. {\bf 3} 351 (1959).
\bibitem{Z1} B. Zumino, J. Math. Phys. {\bf 1} 1 (1960).
\bibitem{W1} J.C. Ward, Phys. Rev. {\bf 78} (1950).
\bibitem{G1} H.S. Green, Proc. Phys. Soc. (London) {\bf A66} 873 (1953).
\bibitem{T1} Y. Takahashi, Nuovo Cimento {\bf 6} 371 (1957).
\bibitem{BR1} C.J. Burden and C.D. Roberts, Phys. Rev. {\bf D47} 5581 (1993).
\bibitem{BC1} J.S. Ball and T.-W. Chiu, Phys. Rev. {\bf D22} 2542
(1980).
\bibitem{H1} B. Haeri, Phys. Rev. {\bf D43} 2701 (1991).
\bibitem{CP1} D.C. Curtis and M.R. Pennington, Phys. Rev. {\bf D42} 4165 
(1990).
\bibitem{dong} Z. Dong, H.J. Munczek and C.D. Roberts,
Phys. Lett. {\bf B333} 536 (1994). 
\bibitem{BP1} A. Bashir and M.R. Pennington, Phys. Rev. {\bf D50} 7679 (1994).
\bibitem{BP2} A. Bashir and M.R. Pennington, Phys. Rev. {\bf D53} 4694 (1996).
\bibitem{BKP1} A. Bashir, A. K{\i}z{\i}lers\"{u} and M.R. Pennington, 
Phys. Rev. {\bf D57} 1242 (1998).
\bibitem{BT1} C.J. Burden and P.C. Tjiang, Phys. Rev {\bf D58} 085019 (1998).
\bibitem{BKP2}  A. Bashir, A. K{\i}z{\i}lers\"{u} and M.R. Pennington, 
Phys. Rev. {\bf D62} 085002 (2000).
\bibitem{davydychev} A.I. Davydychev, P. Osland and L. Saks,
Phys. Rev. {\bf D63} 014022 (2001). 
\bibitem{Ross} E.G. Floratos, D.A. Ross and C.T. Sachrajda,
Nucl. Phys. {\bf B129} 66 (1977).
\bibitem{BKP3} {\em ``Analytic Form of the One Loop Vertex and the Two
Loop Fermion Propagator in 3-Dimensional Massless QED''} by A. Bashir, 
A. K{\i}z{\i}lers\"{u} and M.R. Pennington,
Adelaide University preprint no. ADP-99-8/T353,  Durham University preprint
no. DTP-99/76, hep-ph/9907418.
\bibitem{adnan1} {\em ``Perturbation Theory Constraints on 
the 3-Point Vertex in massless QED3''} by A. Bashir,
Proceedings of
the Workshop on Light-Cone QCD 
and Non Perturbative Hadron
Physics, World Scientific, University of Adelaide, Adelaide, Australia, 
(227-232) 2000.
\bibitem{adnan2} A. Bashir, Phys. Lett. {\bf B491} 280 (2000).
\bibitem{Moch} S. Moch, P. Uwer and S. Weinzierl, J. Math. Phys. {\bf
43} 3363 (2002).
\bibitem{Reenders} A. K{\i}z{\i}lers\"{u}, M. Reenders and
M.R. Pennington, Phys. Rev. {\bf D52} 1242 (1995).
\bibitem{BR1} A. Bashir and A. Raya, Phys. Rev. {\bf D64} 105001 (2001).
\bibitem{Zsums1} L. Euler, Novi Comm. Acad. Sci. Petropol. {\bf 20}
140 (1775).
\bibitem{Zsums2} D. Zagier, First European Congress of Mathematics,
Vol. II, Birkhauser, Boston, 497 (1994).
\bibitem{tables} I.S. Gradshteyn and I.M. Ryzhik, {\em Table of
Integrals, Series and Products, sixth edition, (Academic Press, USA), 2000.}
\bibitem{moretables} M. Abramowitz and I.A. Stegun, 
{\em Handbook of Mathematical Functions, (Dover Publications, USA), 1972.}
\end{thebibliography}
\end{document}